\documentclass[12pt]{article}
\usepackage{graphicx}
\usepackage{amssymb}
\usepackage{graphics}
\usepackage{amsmath}
\usepackage{amsfonts}
\usepackage{bm}

\def\non{\nonumber}

\def\1{{_{1}}}\def\2{{_{2}}}

\def\noHe0{:\;\!\!\;\!\!:H_e(0):\;\!\!\;\!\!:}
\def\noHm0{:\;\!\!\;\!\!:H_\mu(0):\;\!\!\;\!\!:}

\def\non{\nonumber}

\def\1{{_{1}}}\def\2{{_{2}}}

\def\be{\begin{equation}}
\def\ee{\end{equation}}

\def\bea{\begin{eqnarray}}
\def\eea{\end{eqnarray}}

\newcommand\pubnumber{NuPhys2018-Capolupo}
\newcommand\pubdate{\today}

\def\napoli{Dipartimento di Fisica E.R. Caianiello, Universita' di Salerno, ITALY \\
INFN - Gruppo Collegato di Salerno, ITALY}
\def\support{\footnote{Work supported by INFN.}}

\def\Title#1{\begin{center} {\Large #1 } \end{center}}
\def\Author#1{\begin{center}{ \sc #1} \end{center}}
\def\Address#1{\begin{center}{ \it #1} \end{center}}

\newcommand\pubblock{\rightline{\begin{tabular}{l} \pubnumber\\
         \pubdate  \end{tabular}}}
\newenvironment{Abstract}{\begin{quotation}  }{\end{quotation}}
\newenvironment{Presented}{\begin{quotation} \begin{center}
             PRESENTED AT\end{center}\bigskip
      \begin{center}\begin{large}}{\end{large}\end{center} \end{quotation}}

\def\beq{\begin{equation}}
\def\eeq#1{\label{#1}\end{equation}}
\def\eeqn{\end{equation}}

\def\beqa{\begin{eqnarray}}
\def\eeqa#1{\label{#1}\end{eqnarray}}
\def\eeqan{\end{eqnarray}}

\let\bar=\overbar

\def\Dslash{\not{\hbox{\kern-4pt $D$}}}
\def\dslash{\not{\hbox{\kern-2pt $\del$}}}

\def\ee{e^+e^-}

\def\msb{{\bar{\ssstyle M \kern -1pt S}}}

\begin{document}
\begin{titlepage}
\pubblock
\vfill
\Title{Total and geometric phases, Majorana and Dirac neutrinos}
\vfill
\Author{Antonio Capolupo\support}
\Address{\napoli}
\vfill
\begin{Abstract}
The analysis of the total and geometric phases generated by the neutrino oscillation shows that these phases for Majorana neutrinos are depending on the representation of the mixing matrix and they are different from those of Dirac neutrinos.
\end{Abstract}
\vfill
\begin{Presented}
NuPhys2018, Prospects in Neutrino Physics\\
Cavendish Conference Centre, London, UK, December 19--21, 2018
\end{Presented}
\vfill
\end{titlepage}
\def\thefootnote{\fnsymbol{footnote}}
\setcounter{footnote}{0}
%


   {\textbf{- Introduction:}} The discovery of the neutrino oscillation \cite{Nakamura1}-\cite{T2K} has definitively shown that the neutrino has a mass.
It remains to determine the nature of the neutrino. Indeed, since the neutrinos are electrically neutral, they can be Dirac particles (fermions different from their antiparticles), or Majorana particles (fermions coinciding with their antiparticles).
 For Dirac neutrinos, the Lagrangian is invariant under $U(1)$ global transformation. Then  the total lepton charge is conserved
 and, in the case of  the mixing of $n$  fields, the Pontecorvo-Maki-Nakagawa-Sakata (PMNS) mixing matrix  has $N_{D}$ physical phases given by $N_{D} = \frac{(n -1)(n-2)}{2}$.
 By contrast, for Majorana neutrinos, the Lagrangian breaks the $U(1)$  symmetry, then processes violating
 the lepton number (such as the neutrinoless double $\beta$ decay) are allowed  and the mixing matrix contains $N_{M}$ physical phases given by  $N_{M} = \frac{n (n - 1) }{2}$. The  $n -1$ extra phases present in the mixing matrix of Majorana neutrinos are called Majorana phases.
 Many representations of  Majorana mixing matrix can be achieved by the rephasing the charged fields in the charged current weak-interaction Lagrangian \cite{Giunti}. For example, in two flavor mixing, $N_{M} = 1$ and we can consider the following  mixing matrices for Majorana neutrinos
  \bea\label{U1}
 U_{1} =
   \left(
            \begin{array}{cc}
             \cos  \theta   &    \sin  \theta\,  e^{i \phi}\\
             -  \sin  \theta  &  \cos  \theta\,  e^{i \phi}\\
            \end{array}
          \right)
           \,,
           \qquad
          or
         \qquad
U_{2} =    \left(
            \begin{array}{cc}
             \cos  \theta   &    \sin  \theta\,  e^{-i \phi}\\
             -  \sin  \theta\,  e^{i \phi} &  \cos  \theta \\
            \end{array}
          \right)
             \,,
          \eea
where $\theta$ is the mixing angle  and $\phi$ is the Majorana phase. This phase can be removed for Dirac neutrinos.
Notice that the Majorana phases
have no effect in neutrino oscillation formulae that are the same for Dirac   and for Majorana neutrinos \cite{Giunti}. Therefore, the oscillation formulae do not allow to reveal the neutrino nature.
However, recently it has been shown that the phenomenon of the decoherence in neutrino evolution can lead to oscillation formulae for Majorana neutrinos different from those of  Dirac neutrinos \cite{Capolupo:2018hrp}.
Moreover it has been shown that quantities such as the Leggett-Garg $K_{3}$ quantity \cite{Richter} and  the  phases generated in the neutrino oscillation \cite{Capolupo:2016idi}  can in principle discriminates between Dirac and Majorana neutrinos.

In this paper, we report the results presented in Ref.\cite{Capolupo:2016idi} and we show that the phases due to the transitions among different flavors states depend  on the mixing matrix $U$ considered. Indeed, different choices of $U$ for Majorana neutrinos generate different values of the total and   geometric phases.
We show these differences by considering the two flavor neutrino mixing case and the   matrix $\textit{U}_{  1}$ and  $\textit{U}_{ 2}$ in Eq.(\ref{U1}). We demonstrate that by using $\textit{U}_{ 2}$, the  values of the total and geometric phases of Majorana neutrinos  are different from those obtained for  Dirac neutrinos, while, by using $\textit{U}_{  1}$, all the phases are independent from $\phi$ and the results for Majorana and for Dirac neutrinos coincide.
Thus, the total and geometric phase provide a tool to determine the choice of  $U$ and to reveal the nature of the neutrino.


 {\textbf{- Neutrino mixing and phases:}} The study of the interferometry has attracted   a great attention in the recent years and new progress on the detection of the geometric phase \cite{Berry:1984jv}    has been made.
The  geometric phase appears  in the evolution of any quantum state
describing a physical system    characterized by a Hamiltonian defined on a parameter space.

Here, we analyze in  particular, the total phase  and the  non--cyclic geometric phase \cite{Mukunda} (which generalizes the Berry phase to the case of not cyclic, not  adiabatic evolution) for neutrinos propagating in vacuum and through the matter. We  use the mixing matrix $U_2$ to show the differences between Majorana and Dirac neutrino. We also compare the results obtained with those achieved by using the matrix $U_1$. We remind that the matter effects are described by replacing  $\Delta m^{2}$ with $\Delta m_{m}^{2} = \Delta m^{2} R_{\pm}$, and $\sin 2\theta $ with $\sin 2\theta_{m} =  \sin 2\theta /R_{\pm}$, where $R_{\pm} = \sqrt{\left(\cos 2\theta  \pm  \frac{2 \sqrt{2 } G_{F} n_{e} E}{\Delta m^{2}}\right)^{2}+ \sin^{2} 2\theta}\,, $  with $+$ for   antineutrinos and $-$ for  neutrinos \cite{MSW1}.
Then, the results obtained for propagation in vacuum and through the matter are formally the same. In the following we consider the propagation in a medium and we use the quantities $\Delta m_{m}^{2}$ and $\sin 2\theta_{m} $.

The geometric phase for a quantum system with state vector   $|\psi(s)\rangle$, is given by
  the difference between the total phase $\Phi^{tot}_{\psi} = \arg \langle \psi(s_{1})| \psi(s_{2} )\rangle$ and the dynamic phase
$\Phi^{dyn}_{\psi} = \Im\int_{s_{1}}^{s_{2}}\langle \psi(s)|\dot{\psi}(s)\rangle d s$, i.e.
$
\Phi^{g }   =   \Phi^{tot}_{\psi}  - \Phi^{dyn}_{\psi},
 $
where  $s$ is a real parameter such that $s \in [s_1, s_2]$, and
the dot denotes the derivative with respect to  $s$.
We find that the geometric phase for electron neutrino is independent on the Majorana phase, therefore it is the same for   Majorana and for Dirac neutrinos. It is given by
\bea\non\label{fase1}
\Phi^{g }_{\nu_{e}}(z)
 & = &   \arg \left[\langle \nu_{e}(0)| \nu_{e}(z)\rangle  \right] - \Im \int_{0}^{z}  \langle  \nu_{e}(z^{\prime})| \dot{\nu}_{e}(z^{\prime})\rangle  d z^{\prime}
 \\  & = & \arg \left[ \cos \left(\frac{\Delta m_{m}^{2} z}{4 E}\right) + i \cos 2\theta_{m} \sin \left(\frac{\Delta m_{m}^{2} z}{4 E}\right) \right]
- \frac{\Delta m_{m}^{2} z}{4 E} \,\cos 2\theta_{m}  \,.
\eea
Similar result is obtained for  muon neutrino, indeed we have $\Phi^{g }_{\nu_{\mu}}(z) = - \Phi^{g }_{\nu_{e}}(z)$.
Let us consider now the following  phases
due to the transitions $\nu_{e}\rightarrow \nu_{\mu}$
and $\nu_{\mu}\rightarrow \nu_{e}$
 \bea  \label{fasemix1}
\Phi_{\nu_{e}\rightarrow \nu_{\mu}}(z)
\!\!& = &\!\!
\arg \left[\langle \nu_{e}(0)| \nu_{\mu}(z)\rangle  \right] - \Im \int_{0}^{z}  \langle  \nu_{e}(z^{\prime})| \dot{\nu}_{\mu}(z^{\prime})\rangle  d z^{\prime}\, ,
\\  \label{fasemix2}
\Phi_{\nu_{\mu}\rightarrow \nu_{e}}(z)
\!\!& = &\!\!
\arg \left[\langle \nu_{\mu}(0)| \nu_{e}(z)\rangle  \right] - \Im \int_{0}^{z}  \langle  \nu_{\mu}(z^{\prime})| \dot{\nu}_{e}(z^{\prime})\rangle  d z^{\prime}\, .
\eea
They are gauge invariant and reparametrization invariant, therefore they are geometric phases.
 By using the Majorana neutrino states obtained by the $U_2$ mixing matrix, we have  $\Phi _{\nu_{e}\rightarrow \nu_{\mu}} \neq \Phi _{\nu_{\mu}\rightarrow \nu_{e}} $, indeed
 \bea
\label{fasemix1a}
\Phi _{\nu_{e}\rightarrow \nu_{\mu}}(z) &  = &  \frac{3\pi}{2} + \phi
+  \left(\frac{\Delta m_{m}^{2}}{4 E}\,\sin 2\theta_{m}\; \cos \phi\; \right) z\, ,
\\
 \label{fasemix2a}
\Phi _{\nu_{\mu}\rightarrow \nu_{e}}(z)
 &  = &   \frac{3\pi}{2} - \phi
+  \left(\frac{\Delta m_{m}^{2}}{4 E} \sin 2\theta_{m}\; \cos \phi\ \right) z\, .
\eea
 Moreover, for the total phases, we have   $\Phi^{tot }_{\nu_{e}\rightarrow \nu_{\mu}} =  \frac{3\pi}{2} + \phi$ and $ \Phi^{tot}_{\nu_{\mu}\rightarrow \nu_{e}} = \frac{3\pi}{2} - \phi $.
On the contrary, for Dirac neutrinos we obtain
  \bea
 \label{fasemixD}
\Phi _{\nu_{e}\rightarrow \nu_{\mu}}(z) = \Phi _{\nu_{\mu}\rightarrow \nu_{e}}(z)
=
\frac{3\pi}{2}
+  \left(\frac{\Delta m_{m}^{2}}{4 E}\,\sin 2\theta_{m}\;  \right) z \,,
\eea
and  the total phases become   $\Phi^{tot }_{\nu_{e}\rightarrow \nu_{\mu}}(z) = \Phi^{tot}_{\nu_{\mu}\rightarrow \nu_{e}}(z)
= \frac{3\pi}{2} $.
Notice that $\Phi_{\nu_{e}\rightarrow \nu_{\mu}}$, $\Phi_{\nu_{\mu}\rightarrow \nu_{e}}$ and the total phases depend on the choice of the mixing matrix.
Indeed, by using the   mixing matrix  $U_{1}$, we derive   the result of Eq.(\ref{fasemixD})   also for Majorana neutrinos.

In Fig.1 we plot  the total and geometric phases for electron neutrino propagating through the matter. We consider  the values of the parameters of  RENO experiment \cite{RENO}:  neutrino energy   $E \in [2 - 8] MeV$,  electron earth density $n_{e } =10^{24} cm^{-3} $, $\Delta m^{2} = 7.6 \times 10^{-3}eV^{2}$ and   distance $z = 100 km$.
\begin{figure}[btp]
\centering
\includegraphics[width=8.0cm]{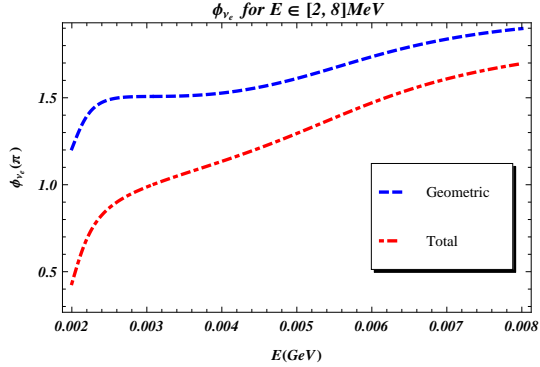}\\
\caption{\em   Plots of the total (the red dot dashed line) and the geometric phases (the blue  dashed line)  of  $ \nu_{e}  $, as a function of   $E$, for a distance length $z = 100 km$.}\label{pdf}
\end{figure}
In Fig.2 we plot    $\Phi _{\nu_{e}\rightarrow \nu_{\mu}} $ and $\Phi _{\nu_{\mu}\rightarrow \nu_{e}} $ presented in Eqs.(\ref{fasemix1a}) and (\ref{fasemix2a}), by considering $\phi = 0.3$ and the values of $n_e$ and $\Delta m^{2}$ of Fig.1. We use the values of the parameters of $T2K$ experiment \cite{T2K}: $E \sim 1 GeV$ and   $z = 300 km$.
\begin{figure}[btp]
\centering
\begin{picture}(300,180)(0,0)
\put(10,20){\resizebox{8.0 cm}{!}{\includegraphics{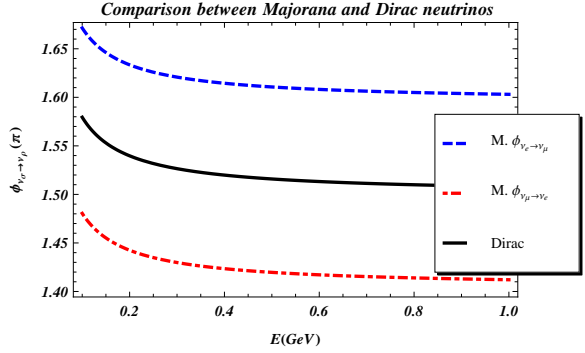}}}
\end{picture}\vspace{-1cm}
\caption{\em   Plot of the  phases $\Phi _{\nu_{e}\rightarrow \nu_{\mu}} $ (the blue dashed line)  and $\Phi _{\nu_{\mu}\rightarrow \nu_{e}} $ (the red dot dashed line) for Majorana neutrinos as a function of   $E$, for   $z = 300 km$. The   phases $\Phi _{\nu_{e}\rightarrow \nu_{\mu}} =\Phi _{\nu_{\mu}\rightarrow \nu_{e}} $ for Dirac neutrinos  is represented by the black solid line.}
\label{pdf}
\end{figure}


 {\textbf{- Conclusions:}} We have shown that, for Majorana neutrinos, the total and the geometric phases due to the  oscillation
assume different values depending on the representation of the mixing matrix.
In particular, we have proved that the total and the geometric phases of Majorana neutrino can be different from those of Dirac neutrino.
Therefore these phases can represent a new tool to study the nature of neutrinos.
We have presented also a numerical analysis and plotted the geometric phases by using  the characteristic parameters of RENO and T2K experiments.

\end{document}